\documentclass[sigconf,9pt,pbalance]{acmart}
\acmConference[DXConf '26]{1st ACM Conference on Digital Transformation}{October 29--30, 2026}{Ann Arbor, MI, USA}
\acmYear{2026}
\copyrightyear{2026}
\IfFileExists{rights.tex}{\copyrightyear{2026}
\acmYear{2026}
\setcopyright{cc}
\setcctype{by}
\acmConference[ACM DXConf 2026]{The Inaugural ACM Conference on Digital Transformation}{October 29--30, 2026}{Ann Arbor, MI, USA}
\acmBooktitle{The Inaugural ACM Conference on Digital Transformation (ACM DXConf 2026), October 29--30, 2026, Ann Arbor, MI, USA}
\acmDOI{10.1145/3847238.3849038}
\acmISBN{979-8-4007-3028-3/2026/10}
}{%
  \setcopyright{none}
  \acmDOI{}
  \acmISBN{}
  \PackageWarningNoLine{camera-ready}{Paper-specific ACM eRights metadata is pending}
}
\usepackage{amsmath}
\usepackage{textcomp}
\usepackage{microtype}
\newcommand{\demourl}{\url{https://tellegen.dev}}
\newcommand{\repourl}{\url{https://github.com/eigenergy/tellegen}}

\newcommand{\im}{\mathrm{j}}

\newif\ifshowedits
\showeditsfalse
\DeclareRobustCommand{\edit}[1]{\ifshowedits\textcolor{blue}{#1}\else#1\fi}

\begin{document}
\title[Interactive Power Flow]{Interactive Power Flow in the Browser}

\author{Samuel Talkington}
\affiliation{%
  \institution{Harvard University}
  \city{Allston}
  \state{MA}
  \country{USA}
}
\affiliation{%
  \institution{University of Michigan}
  \city{Ann Arbor}
  \state{MI}
  \country{USA}
}
\email{talks@umich.edu}

\author{Frederik Geth}
\affiliation{%
  \institution{The University of Queensland}
  \city{Springfield}
  \state{QLD}
  \country{Australia}
}
\email{f.geth@uq.edu.au}

\author{Qian Zhang}
\affiliation{%
  \institution{Harvard University}
  \city{Allston}
  \state{MA}
  \country{USA}
}
\email{qianzhang@g.harvard.edu}

\author{Le Xie}
\affiliation{%
  \institution{Harvard University}
  \city{Allston}
  \state{MA}
  \country{USA}
}
\email{xie@seas.harvard.edu}

\author{Skyler Liu}
\affiliation{%
  \institution{Harvard University}
  \city{Allston}
  \state{MA}
  \country{USA}
}
\email{skliu@g.harvard.edu}

\renewcommand{\shortauthors}{Talkington, Geth, Zhang, Xie, and Liu}

\begin{abstract}
This paper introduces \texttt{tellegen}, an open source framework for interactive power flow (PF) and optimal power flow (OPF) studies that run in the browser. This provides intuitive and democratized access to power system analysis tools compiled to WebAssembly. A user can drag and drop a case file, click and drag to change a nodal demand or line rating, preview the impacts via sensitivity analysis, and obtain an exact re-solve on release. User case files and results stay entirely on the device: \texttt{tellegen} transmits \textit{zero} Critical Energy/Electric Infrastructure Information (CEII). The framework comprises a core numerical engine for PF and OPF, reusable browser components, saved studies, and structured WebMCP tools for agentic interaction. We evaluate the transmission OPF solver by comparing objectives with PGLib baselines; the distribution PF solver by comparing voltages and currents with OpenDSS; and the WebAssembly execution times by comparing with PowerModels.jl. \edit{On realistic synthetic grids, WebAssembly OPF solves take only 25--43\% longer than native binary solves.} The implementation shows how an engineer can distribute an executable numerical study as a URL, reducing installation and hosting requirements while keeping case data local.
\end{abstract}


\maketitle
\hypersetup{pdfauthor={Samuel Talkington, Frederik Geth, Qian Zhang, Le Xie, Skyler Liu}}

\section{Introduction}\label{sec:intro}
This paper presents a web development toolbox that allows interactive sharing of a power system study. With current power system analysis software, it is straightforward to generate static illustrations of voltages or power flows for presentation to an audience. However, changing the assumptions of the study usually requires installing numerical software, obtaining its dependencies, and learning its input conventions. Open source tools such as MATPOWER~\cite{zimmerman2011matpower}, pandapower~\cite{thurner2018pandapower}, and PowerModels.jl~\cite{coffrin2018powermodels} remove license barriers and provide established methods. Delivering a focused, easily executable study to someone without access to the author's environment remains a separate problem.

The \texttt{tellegen} framework addresses this delivery problem by empowering power engineering practitioners, students, researchers, and the general public to distribute an electric power system study as a website link. Any recipient of the link can open the study, change demand, a line rating, or another electrical parameter, and then visually inspect the impacts via sensitivity analyses. Furthermore, a new solution to the study can be obtained directly in the browser when edits are committed. A \texttt{tellegen} application can also accept local case files through an intuitive drag-and-drop operation. Parsing and solving the case file happens entirely on the user's device, with no case upload or remote solver required. This makes it possible to develop a broad array of web applications useful for public outreach, for teaching examples, and for exploration of sensitive network models.

The numerical engine of \texttt{tellegen} is written in Rust and compiled to WebAssembly (\texttt{wasm}), a portable binary instruction format executed by modern web browsers~\cite{haas2017wasm}. It currently includes capabilities for transmission power flow (PF) and optimal power flow (OPF) formulations and multiconductor distribution PF. The OPF interface supports first-order sensitivity analysis via implicit differentiation of the OPF solution, with exact re-solves, all happening entirely in the browser. The PF interface uses retained numerical sessions for repeated load changes. A separate library, \texttt{powerio}, supplies case parsing, network representations, and problem assembly. A browser component library is also provided to connect these calculations to maps, equipment inspection, and offline saving of studies. In addition, we provide experimental support for the emerging WebMCP standard, which exposes structured requests for a browser-native AI agent to inspect and operate the same study displayed to the human user.

\paragraph{Contribution}
We present an open source framework for interactive power system analysis and visualization in the browser. For the transmission capabilities, we evaluate the performance of the OPF solver against PowerModels.jl as a baseline. For the distribution capabilities, we compare PF solutions against OpenDSS as a baseline. The primary contribution of our work is the integrated web framework and its evaluation against existing tools. This work provides a proof of concept demonstration of how local numerical execution and reusable data models make an industrial analysis accessible through an ordinary web link.

A demo application (\demourl) showcases the capabilities of the framework, which includes four synthetic transmission grids and a distribution example. The rest of the paper is organized as follows. Section~\ref{sec:architecture} introduces its models and interaction design; Sections~\ref{sec:eval} and~\ref{sec:demo} present evaluation and use; Section~\ref{sec:discussion} discusses access, limitations, and lessons for other numerical applications. The source and browser packages are available at \repourl~\cite{tellegen2026repo}.

\begin{figure*}[t]
  \centering
  \includegraphics[width=\textwidth]{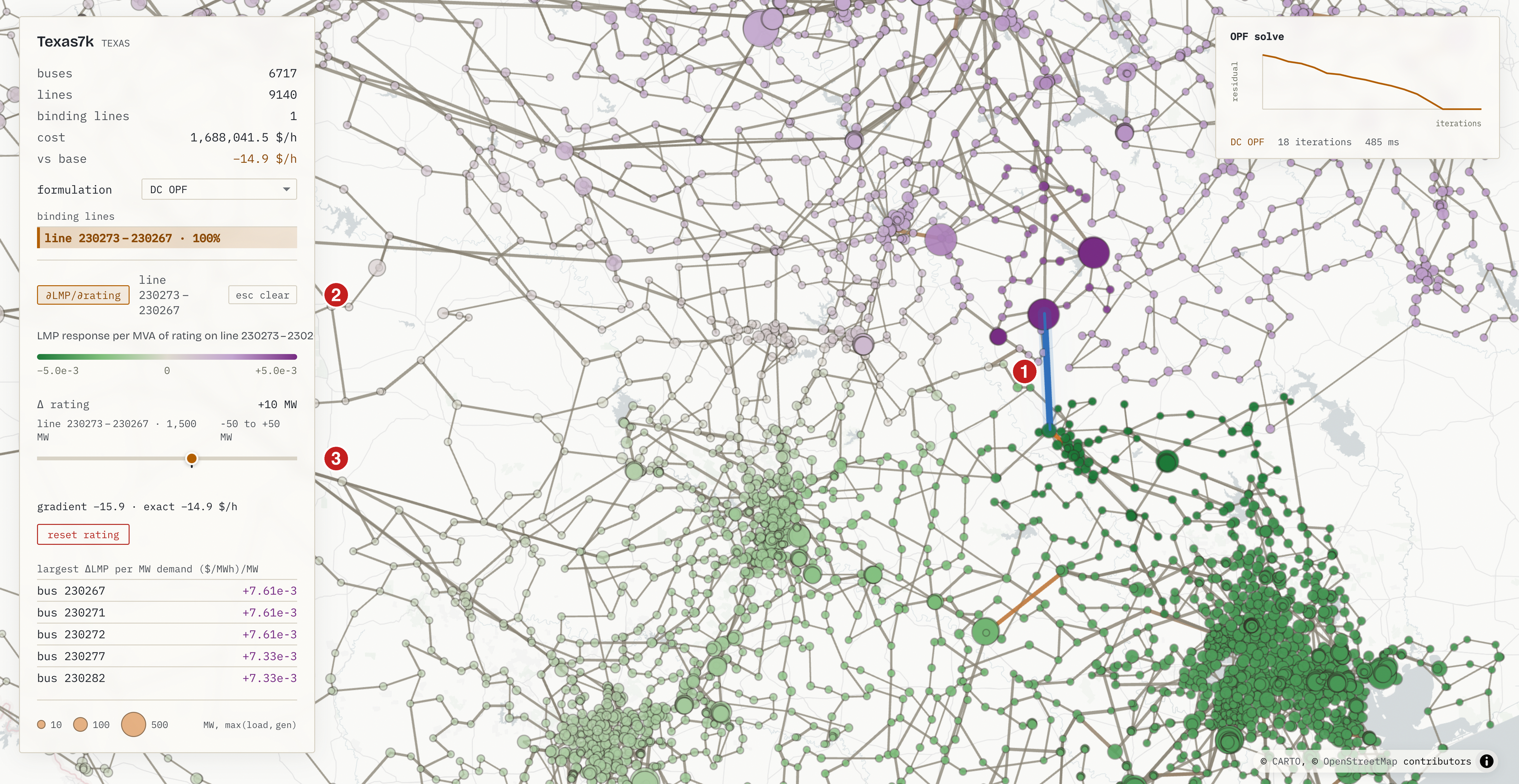}
  \caption{A Texas7k study after a $+10$~MW \edit{line rating} edit: (1) a binding line \edit{$\ell$}; (2) the sensitivity \edit{$\partial\lambda_i/\partial\bar f_\ell$} across buses \edit{$i$}; (3) the rating control and predicted cost change ($-15.9$~\$/h), compared with the re-solved change ($-14.9$~\$/h).}
  \Description{A map of the synthetic Texas network with a selected binding line and buses colored by the sign and magnitude of their price sensitivity to that line's rating. A side panel shows the rating edit and compares its predicted and re-solved cost changes.}
  \label{fig:teaser}
\end{figure*}
\section{System Architecture}\label{sec:architecture}
\begin{figure}[t]
\centering
\includegraphics[width=\linewidth]{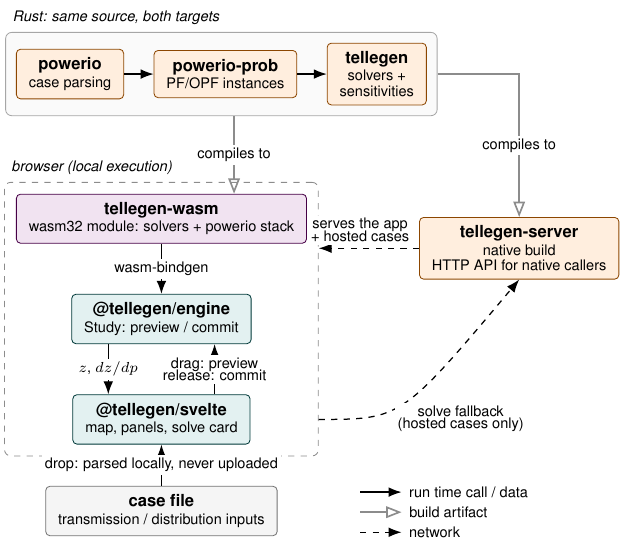}
\caption{Architecture of the \edit{\texttt{tellegen}} framework. Local case files are parsed and used to build and solve problem instances directly in the browser. An optional server library can be used to host public cases.}
\Description{A dependency diagram connects powerio parsing and problem assembly to the tellegen Rust engine. Build arrows lead to a browser WebAssembly module and a native server. Inside the browser, an engine package connects the solver to interactive map components; a local file enters this path without going to the server. Dashed arrows indicate application delivery and an optional server fallback for public hosted cases.}
\label{fig:architecture}
\end{figure}

The key idea behind \texttt{tellegen} is combining a compiled OPF solver with graphical components for loading, editing, and displaying power system models. The underlying OPF solver can be compiled for both native (binary) and browser (\texttt{wasm}) execution. \edit{With the solver compiled to \texttt{wasm}}, the browser performs case parsing and numerical computation locally, so a study can be distributed as a web application without operating a solver service on a cloud server. Fig.~\ref{fig:architecture} describes the architecture of the \texttt{tellegen} framework.

\subsection{Models and formulations}
Power flow (PF) computes a network's steady state for specified injections and voltage controls. Optimal power flow (OPF) chooses generator setpoints or other controllable quantities to minimize total operational costs subject to the power flow equations and engineering constraints. Transmission studies commonly represent a balanced three-phase system by one \edit{positive sequence} network. Distribution studies retain individual conductors to represent phase imbalance and neutral displacement. Table~\ref{tab:forms} separates these physical representations from the mathematical problems and their numerical methods.

\begin{table*}[t]
\centering
\caption{Formulations in the transmission and distribution stack. PF specifies operating conditions; OPF optimizes them. LinDist3Flow uses linearized network equations, with conic limits where selected.}
\label{tab:forms}
\begin{tabular}{lllll}
\toprule
Network representation & Task & Formulation & Numerical method & Availability \\
\midrule
Positive sequence & PF & DC approximation & Sparse linear solve & Native and browser \\
Positive sequence & PF & Nonlinear AC, polar voltages & Newton--Raphson & Native and browser \\
Positive sequence & OPF & DC, convex quadratic costs & Clarabel (QP) & Native and browser \\
Positive sequence & OPF & SOCWR relaxation of AC OPF & Clarabel (conic) & Native and browser \\
Individual conductors & PF & Nonlinear AC, complex voltages & Fixed-point current injection & Native and browser \\
Phase conductors & PF/OPF & LinDist3Flow approximation & Affine/conic program & \edit{\texttt{powerio}} assembly; UI pending \\
\bottomrule
\end{tabular}
\end{table*}

\paragraph{DC PF and OPF}
The DC approximation treats active power flow in an \edit{alternating current} (AC) network as a lossless linear problem. Let $\theta$ be a vector of bus voltage angles, $g$ a vector of active power generation, and $d$ a vector of active power demand. Let $C$ denote the generator-to-bus incidence matrix. With a reference angle fixed, DC PF solves
\begin{equation}
 B\theta=Cg-d-h,\qquad f=H\theta+\edit{f_0},\label{eq:dc}
\end{equation}
where $B$ and $H$ are nodal and branch susceptance matrices, respectively, and $f$ is a vector of branch active power flows. The vectors $h,\edit{f_0}$ account for fixed shunts and phase shifts; we follow the conventions of MATPOWER in this regard, see~\cite{zimmerman2011matpower}.

DC OPF minimizes an operational cost function, e.g., $\sum_k(a_k g_k^2+b_kg_k+c_k)$, with $a_k\geq0$, subject to~\eqref{eq:dc}, generator limits, line thermal limits \edit{$|f_\ell|\leq\bar f_\ell$}, and phase angle difference limits. The Lagrange multipliers associated with the \edit{nodal power balance equations}, denoted as~$\lambda$, are known as locational marginal prices (LMPs). The LMP~$\lambda_i$ describes the change in \edit{optimal} operational cost with respect to \edit{demand $d_i$ at bus $i$}.

\paragraph{AC PF, AC OPF, and the SOC relaxation}
For a vector of \edit{$n$} complex bus voltages $u$ and bus admittance matrix $Y$, the AC power flow equations are
\begin{equation}
 s_i =u_i\overline{(Yu)_i}, \quad i=1,\ldots,n\label{eq:ac}
\end{equation}
where \edit{$s_i=p_i+\im q_i$} is the net complex power injection \edit{at bus $i$}. ``PQ buses'' specify their active and reactive power injection; ``PV buses'' specify their active power injection and voltage magnitude. In transmission systems, it is common to apply the Newton-Raphson method to solve the active and reactive power mismatches at PQ buses and the active mismatches at PV buses, with a prescribed reference voltage.

AC OPF seeks to minimize an operational cost function subject to the AC power flow equations~\eqref{eq:ac} and engineering constraints. For more information on this \edit{nonconvex} optimization problem, see~\cite{molzahn_hiskens-fnt2019}. For the purposes of \texttt{tellegen}, we focus on convex relaxations of the AC OPF problem, as mature convex optimization solvers readily compile to \texttt{wasm}. The second-order cone relaxation of AC OPF in \edit{voltage product} variables (SOCWR) lifts voltage products to $w_i:=|u_i|^2$ and $W_{ij}=c_{ij}+\edit{\im}s_{ij}=u_i\overline{u_j}$. Nodal power injections and branch power flows become affine in these variables. The relaxation imposes
\begin{equation}
 c_{ij}^2+s_{ij}^2\leq w_iw_j,\qquad w_i,w_j\geq0,\label{eq:soc}
\end{equation}
alongside nodal power balance and limits on generation, squared voltage magnitudes, apparent power, and other desired constraints~\cite{jabr2006socwr,coffrin2018powermodels}. Minimizing convex generation cost gives a convex optimization problem readily solved by Clarabel~\cite{goulart2024clarabel}, though a relaxed solution need not admit globally consistent AC voltage phasors~\cite{low2014relaxation}. Our implementation of the SOC relaxation of AC OPF has the benefit of requiring no external BLAS library, which at the time of writing was a required dependency for general nonlinear programming solvers needed for the full AC OPF problem. This enables \texttt{tellegen} to easily compile to \texttt{wasm}.

\paragraph{Multiconductor AC PF}
The distribution solver uses a compensated \edit{current injection} iteration of the type used in OpenDSS~\cite{dugan2011opendss}. With \edit{subscripts $r$ and $\ell$ denoting prescribed reference terminals and unknown terminals, respectively, iteration $k$ solves}
\begin{equation}
 \edit{\widehat Y_{\ell\ell}u_{\ell}^{(k+1)}=\iota_{\ell}\big(u^{(k)}\big)-\widehat Y_{\ell r}u_r},\label{eq:mcpf}
\end{equation}
where \edit{$u$ contains terminal voltage phasors and $\widehat Y$} combines the passive network with fixed reference load admittances. \edit{The compensation currents $\iota_\ell$ at unknown terminals} subtract the nonlinear load withdrawals and add back their \edit{reference admittance} currents. One sparse LU factorization serves the voltage iterations. The model retains explicit neutrals, line coupling, shunts, supported winding connections, and \edit{constant power}, \edit{constant current}, \edit{constant impedance}, ZIP, and exponential loads. Distribution extensions add phase-to-phase and center-tap transformers and retained sessions for load edits: changing load powers reuses the prepared matrix and starts from the previous converged voltages. Changes to topology, taps, or the reference source require rebuilding the session.

\paragraph{LinDist3Flow}
This distribution approximation~\cite{ganlowmultiphase} uses linear power balance and squared voltage drop equations with phase coupling, plus linear or second-order cone operating limits. OPF minimizes the selected cost. The PF implementation prescribes injections and reports thermal overloads instead of enforcing those thermal limits. Importantly, our preparation step records component approximations\edit{. For example, the neutral is eliminated using} a Schur complement under \edit{an assumption of zero voltage drop along the neutral}. The meshed extension retains voltage drop and nodal balance equations but adds no angle recovery or loop flow constraints. Its solution is naturally an approximate model and is not an AC feasibility certificate.

\subsection{Sensitivities, preview, and re-solve}
For DC OPF, transmission AC PF, and SOCWR, a common differentiation interface maps supported inputs (demands, costs, limits, and network parameters) to supported outputs (prices, dispatch, flows, and voltages). PF results do not have OPF prices, and the multiconductor fixed-point solver uses repeated solves rather than this sensitivity interface.

Let $K(z,p)=0$ collect the power flow equations or OPF optimality conditions, where~$z$ denotes the solution variables and~$p$ denotes the problem parameters. If~$K$ is continuously differentiable and~$K_z:=\partial K/\partial z$ is nonsingular at a solution, the implicit function theorem gives
\begin{equation}
 K_z\frac{dz}{dp}=-K_p.\label{eq:sens}
\end{equation}
\edit{Here $K_p:=\partial K/\partial p$, and both Jacobians are evaluated at the solution.}
For a smooth constrained optimization problem, constraint independence, strict complementarity, and second-order sufficiency provide familiar sufficient conditions~\cite{fiacco1976sensitivity}; conic degeneracy and changes in the active constraints require particular care. The implementation uses faer~\cite{faer2026} for forward or adjoint linear solves, choosing the direction from the requested input and output dimensions. A factorization serves the right-hand sides within a sensitivity request. This is the same \edit{implicit differentiation} principle used in optimization layers~\cite{amos2017optnet,agrawal2019cvxpylayers}.

A stateful study object retains the solved model. During a slider drag, a preview applies the local first-order response; releasing the control commits the edit and solves the model again. Throughout this paper, an \emph{exact re-solve} means solving the selected formulation to numerical tolerance rather than extrapolating its derivative. It does not imply exact arithmetic, an AC-feasible SOC relaxation, or a globally optimal nonlinear solution.

\paragraph{Infeasibility and numerical failure}
The Clarabel driver distinguishes reported primal infeasibility, dual infeasibility, and other unsuccessful terminations, including numerical errors and iteration limits. It accepts both \texttt{Solved} and \texttt{AlmostSolved}; the latter indicates reduced solver tolerances. Newton and fixed-point methods can fail to converge without proving infeasibility. The multiconductor solver checks both voltage updates and physical KCL residuals. Failed interactive solves report an error and retain the previous committed result; worker errors and unavailable sensitivities are also presented in the UI.

Near singular operating points, derivatives may become large or cease to exist. The OPF sensitivity paths regularize their linear systems and apply iterative refinement, but do not currently report a condition estimate or a certified \edit{derivative error} bound. Factorization errors are reported; successful factorization alone does not establish reliable sensitivities. The preview/re-solve comparison exposes approximation error for completed edits. Crossing an active-set boundary or leaving the feasible region can invalidate a preview, so it must not be treated as a solved operating point.

\subsection{Case data and browser packages}
\texttt{powerio}~\cite{powerio2026repo} owns file parsing, typed network models, calculation instances, portable solutions, and reusable matrix assembly. \edit{\texttt{tellegen}} supplies numerical execution. An AC OPF instance can thus feed the SOC relaxation without confusing its output with a full AC OPF solution. The browser adapter, \texttt{@tellegen/engine}, provides worker execution and the study lifecycle; \texttt{@tellegen/svelte} supplies the map, panels, file loading, and result displays. Rendering uses JavaScript components, deck.gl, and MapLibre, rather than Rust.

MATPOWER \texttt{.m}, PowerWorld \texttt{.aux}, and PSS/E \texttt{.raw} files are parsed locally. Companion display files supply geometry separately from electrical data. Distribution import includes OpenDSS, PowerModelsDistribution engineering JSON~\cite{pmd}, and BMOPF JSON; parsing a format does not guarantee that a solver can operate on the case study data. Multiconductor calculation uses supported BMOPF or typed \edit{\texttt{powerio}} inputs. Local files have no \edit{fallback to solving on a server}. Their contents and electrical results remain on the device: \edit{\texttt{tellegen}} transmits zero Critical Energy/Electric Infrastructure Information (CEII) from these studies. \edit{Server APIs for hosted cases} are separate and operate on the public demonstration data.

\subsection{Access outside the browser}

The same numerical engine is available through several native interfaces outside the browser. Users can invoke the Rust crate directly in native applications, while a Python package exposes the same functionality through PyO3 and is distributed as a binary wheel built with maturin. The latter allows the compiled solvers to run in scripts and notebooks without requiring users to install or manage a Rust toolchain. The interface preserves the same study lifecycle used by the browser: a case is loaded from a file or a \edit{\texttt{powerio}} intermediate representation, a formulation is selected, demand or rating edits are applied and \edit{the model is re-solved}, supported sensitivities are requested, and the base state can be restored. Responses include only the result blocks defined by the selected formulation. A study developed interactively can therefore be continued in batch without introducing separate solver semantics.

The same Python binding also supports a Model Context Protocol (MCP) server~\cite{mcp2025}, exposing these calculations to \edit{language model} agents. The \texttt{tellegen} MCP server is also available through PowerMCP \cite{zhang2025poweragent}. This server complements the browser-facing WebMCP interface described \edit{in Section~\ref{sec:demo}}. WebMCP operates on the study currently displayed in a browser tab, whereas the MCP server reads and writes study files, making it suitable for headless, scripted, and batch workflows. The MCP server has no user interface in which an operator can approve a capacity proposal, but studies exported from either interface can be imported into the other, carrying their states, goals, and supporting evidence.

\section{Evaluation}\label{sec:eval}

\begin{table}[ht]
\caption{Synthetic grids served by the hosted reference demo}
\begin{center}
\begin{tabular}{llrr}
\hline
\textbf{Case} & \textbf{Territory} & \textbf{Buses} & \textbf{Branches} \\
\hline
\texttt{ACTIVSg200}~\cite{birchfield2017activsg} & central Illinois & 200 & 245 \\
\texttt{ACTIVSg500}~\cite{birchfield2017activsg} & South Carolina & 500 & 597 \\
\texttt{Texas7k}~\cite{birchfield2017activsg} & Texas & 6717 & 9140 \\
\texttt{CATS}~\cite{taylor2024cats} & California & 8870 & 10823 \\
\hline
\end{tabular}
\label{tab:cases}
\end{center}
\end{table}

The four transmission cases in the hosted demo (\demourl) comprise three Texas A\&M University synthetic grids~\cite{birchfield2017activsg} and the CATS California Test System~\cite{taylor2024cats} synthetic grid, summarized in Table~\ref{tab:cases}. These synthetic networks range from 200 to 8870 buses and represent geographic footprints without reproducing the corresponding utility network models.

\subsection{Transmission objectives and sensitivities}
Our framework was evaluated against PGLib~\cite{babaeinejad2019pglib}, covering 66 cases at each of three operating conditions: typical, heavily loaded (\texttt{api}), and small angle difference (\texttt{sad}). The 198 attempted case--variant pairs range from 3 to 78,484 buses; we successfully completed solves for cases up to 30,000 buses in the browser. Results by case are provided in the repository documentation~\cite{tellegen2026repo}.

For DC OPF, 148 objectives agree with PowerModels.jl+Ipopt within 1\%, with median relative discrepancy $1.0\times10^{-5}$. Another 45 cases are infeasible in both implementations. Two cases failed numerically, and three exceeded the 180~s budget. For the SOC relaxation, 189 computed objectives lie below the AC OPF reference objectives within tolerance. Of these, 183 reproduce the SOC gap, and their median absolute gap discrepancy is 0.006 percentage points. Six have looser gaps, six fail numerically, and three exceed the time limit. Certifying these objectives as lower bounds would require a feasible relaxation dual. Likewise, a local AC OPF solution need not be the global optimum.

While we have validated objective values obtained by our solvers, we note that solutions may differ when optima are nonunique. We validate sensitivities by comparing them against finite difference estimates. As finite differences are an approximation of the actual derivatives, this is an inherent limitation that requires future work. Changes in the active set, small derivatives, and conic degeneracy may require examining the operating point and error scale.

\subsection{Distribution power flow against OpenDSS}
We compare complex terminal voltages and physical currents obtained by our multiconductor AC PF solver against OpenDSSDirect.py~0.9.4 with DSS C-API~0.14.5. Table~\ref{tab:tellegen-opendss-qc} summarizes the comparisons by feeder. Here $\edit{\Delta u}$ is the difference in the obtained  voltage phasors from OpenDSS, and $r_I$ is \edit{\texttt{tellegen}}'s \edit{current balance} residual; the latter measures satisfaction of Kirchhoff's current law (KCL).

\begin{table}[t]
\centering
\caption{\edit{Distribution PF comparison with OpenDSS using matched source voltages.} Errors and residuals are maxima over compared terminals.}
\label{tab:tellegen-opendss-qc}
\begin{tabular}{lrrr}
\toprule
System & Terminals & $|\edit{\Delta u}|$ (V) & $|r_I|$ (A) \\
\midrule
ENWL\textsuperscript{a} & $\leq28{,}092$ & $1.17\times10^{-7}$ & $4.45\times10^{-7}$ \\
Australian MV & 984 & $4.68\times10^{-6}$ & $2.28\times10^{-7}$ \\
IEEE 13 & 41 & $2.88\times10^{-2}$ & $1.19\times10^{-8}$ \\
IEEE 34 & 138 & $1.97\times10^{-3}$ & $7.68\times10^{-9}$ \\
IEEE 123\textsuperscript{b} & 271 & $8.73\times10^{-3}$ & $9.23\times10^{-7}$ \\
\bottomrule
\end{tabular}
\begin{minipage}{\linewidth}
\footnotesize
\textsuperscript{a}127 of 128 feeders converged; the remaining feeder also failed to solve in OpenDSS after 1,000 iterations. The terminal count is the largest compared feeder.\\
\textsuperscript{b}The export's disconnected bus 610 was removed. IEEE counts include compared phase nodes only.
\end{minipage}
\end{table}

The ENWL networks use the four-wire dataset~\cite{heidari2024fourwire}; the Australian feeder and IEEE feeders are described in~\cite{geth2025realistic,schneider2018analytic}. These comparisons match \edit{\texttt{tellegen}}'s prescribed source voltages by replacing finite OpenDSS source impedances with near-ideal sources and disabling active controls. IEEE~34 retains six fixed regulator phases. The IEEE~123 export omits an unloaded transformer but retains its isolated secondary bus; the unmodified export is singular. The Australian MV case represents closed switches as \edit{finite impedance} lines and adds $20\,\mathrm{kW}+\edit{\im}6\,\mathrm{kvar}$ delta loads at 43 leaf buses.

Separate line/load port comparisons give maximum current discrepancies of $3.80\times10^{-8}$~A on ENWL network~1 feeder~1 and $1.26\times10^{-6}$~A on the Australian feeder. Nine small networks compare terminal voltages, line/load currents and powers, and \edit{total complex power balance between the source and network elements}. Transformer and reactor ports in that nine-case suite are checked for presence and finite values; transformer accuracy is also tested independently through 21 \edit{admittance matrix} comparisons and loaded wye--wye and delta--delta cases. A further 56 reference cases exercise \edit{constant power}, \edit{constant current}, \edit{constant impedance}, ZIP, and exponential loads, with maximum voltage and current discrepancies of $1.08\times10^{-7}$~V and $6.78\times10^{-9}$~A. This evidence establishes agreement for the specified static models; it does not test \edit{time series} controls or every OpenDSS device.

\subsection{LinDist3Flow validation}
The LinDist3Flow checks cover network preparation, phase coupling, power balance, voltage drops, conic limits, and SI/per-unit equivalence. \edit{An analytical feeder with an explicit neutral} checks the neutral reduction and assembled constraints against hand-derived values. An independent nonlinear OpenDSS solve gives a receiving voltage of 220.1931~V, compared with 220.4269~V for the lossless linear model, a 0.1062\% magnitude discrepancy. This is an \edit{approximation error} measurement on one feeder, distinct from the much smaller numerical discrepancies between implementations of the same nonlinear model in Table~\ref{tab:tellegen-opendss-qc}.

\subsection{Transmission OPF solve time}
\begin{table}[ht]
\caption{\edit{Native binary and browser wasm DC OPF solve time (median, ms)}}
\begin{center}
\small
\begin{tabular}{lrr}
\hline
\textbf{Case} & Native & Wasm \\
\hline
\texttt{ACTIVSg200} & 4.0 & 5 \\
\texttt{ACTIVSg500} & 11.7 & 17 \\
\texttt{Texas7k} & 230 & 328 \\
\texttt{CATS} & 388 & 540 \\
\hline
\end{tabular}
\label{tab:timing-dc}
\end{center}
\end{table}

\begin{table}[ht]
\caption{\edit{Native binary and browser wasm SOC relaxation solve time (median, ms)}}
\begin{center}
\small
\begin{tabular}{lrr}
\hline
\textbf{Case} & Native & Wasm \\
\hline
\texttt{ACTIVSg200} & 25.1 & 32 \\
\texttt{ACTIVSg500} & 82 & 103 \\
\texttt{Texas7k} & 1687 & 2199 \\
\texttt{CATS} & 2588 & 3424 \\
\hline
\end{tabular}
\label{tab:timing-socwr}
\end{center}
\end{table}

Tables~\ref{tab:timing-dc} and~\ref{tab:timing-socwr} \edit{compare native and browser solve times on realistic synthetic grids}. Both runs time repeated small edits to load demand parameters, with the model constructed before timing. The browser uses a module worker in Chromium. \edit{Browser solves take only approximately 25--43\% longer than native solves across these cases, based on the unrounded medians.}

\begin{figure}[htbp]
\centerline{\includegraphics[width=\linewidth]{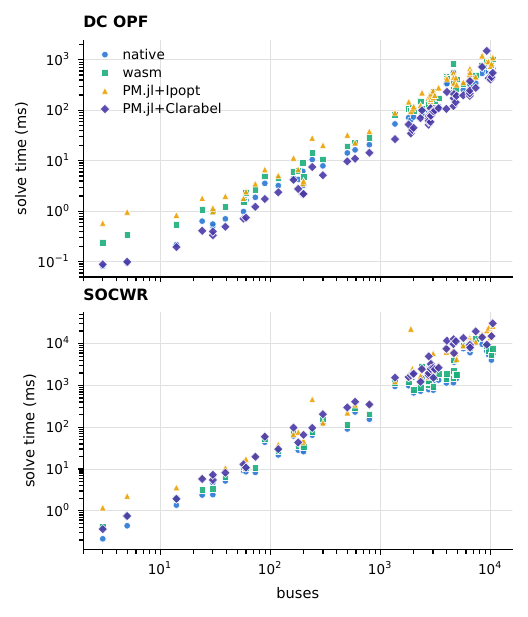}}
\caption{Total solve time vs. bus count on 60 PGLib cases at the default operating point: DC OPF (top) and the SOC relaxation (bottom). \emph{native}: binary solver executed locally; \emph{wasm}: wasm solver executed in the browser; \emph{PM.jl+X}: PowerModels.jl, excluding JuMP model construction time, executed with the specified optimizer.}
\Description{Two stacked log-log scatter plots of solve time in milliseconds against bus count, one for DC OPF and one for SOCWR, each with four series: native, wasm, PowerModels.jl with Ipopt, and PowerModels.jl with Clarabel. Solve times generally increase with bus count. Distinct marker shapes identify the four implementations; the panels use the same bus count axis.}
\label{fig:scaling}
\end{figure}

Fig.~\ref{fig:scaling} extends the comparison to 60 typical PGLib cases through 10,480 buses. The benchmark records native solve time per case, including formulation assembly from a prepared network; wasm reports browser-based solve times, and the PowerModels measurements are optimizer-reported times that exclude JuMP model construction. Median wasm/native ratios computed from the retained samples are 1.43 for DC OPF and 1.25 for SOCWR. PowerModels with Clarabel has lower DC optimizer time than \edit{\texttt{tellegen}}'s native solve stage on the median matched case. For SOCWR, median PowerModels/native ratios are 2.17 with Ipopt and 1.69 with Clarabel.

The 10,000-bus GOC case fails SOCWR numerically in native \edit{\texttt{tellegen}}, wasm, and PowerModels with Clarabel. Three additional PowerModels/Clarabel SOCWR runs fail (197-bus SNEM, 4,917-bus GOC, and 9,591-bus GOC); their points are omitted. Fifteen further PowerModels/Clarabel SOCWR results have \texttt{ALMOST\_OPTIMAL} status and are plotted at their reported times. These failures and approximate terminations remain part of the evaluation.

\subsection{Native distribution PF time}
Table~\ref{tab:mc-pf-wall-time} reports native distribution timings measured on a 16-core Apple M4 Max with 48~GB memory under macOS~26.6.2. Each case had 10 warmups and 100 interleaved measurements with randomized feeder order. Times include fresh-process startup, parsing, preparation, solution, and JSON serialization. The median launch-only baseline was 2.70~ms. Each solve used one LU factorization, so the table reports iterations and \edit{sparse system} size rather than repeating that constant. These measurements characterize native end-to-end latency; they are neither browser timings nor an OpenDSS runtime comparison.

\begin{table}[htbp]
    \centering
    \small
    \setlength{\tabcolsep}{5pt}
    \renewcommand{\arraystretch}{1.1}
    \caption{Native distribution PF latency, including process startup. $n$ is the number of unknown terminals; Nonz. counts matrix nonzeros, and It. counts fixed-point iterations.}
    \label{tab:mc-pf-wall-time}
    \begin{tabular}{lrrrc}
        \hline
        Feeder & \(n\) & Nonz. & It. &
        Median [\edit{p95}] (ms) \\
        \hline
        IEEE 13                  &   38 &    246 & 13 & \(5.67\ [6.21]\) \\
        IEEE 34                  &  135 &  1,071 & 15 & \(6.91\ [7.47]\) \\
        IEEE 123 (modified)       &  268 &  1,928 & 12 & \(9.16\ [9.83]\) \\
        Australian MV        &  981 &  7,803 & 5 & \(14.96\ [15.66]\) \\
        \hline
    \end{tabular}
\end{table}

\section{Interactive Studies}\label{sec:demo}

The browser workflow follows Fig.~\ref{fig:architecture}: \texttt{powerio} parses the file into a canonical network; the \edit{same Rust engine is compiled} to a native target and to \texttt{wasm32}; \texttt{@tellegen/engine} loads the \texttt{wasm} module and exposes the \texttt{Study} lifecycle; and \texttt{@tellegen/svelte} renders the result as a map and a solve card. A formulation selector for transmission networks switches the active solve between DC OPF and the SOCWR relaxation without leaving the page.

In the demo, bus color by default encodes the LMP at the current operating point. Selecting \edit{bus $j$ displays $\partial\lambda_i/\partial d_j$}; selecting \edit{a binding line $\ell$ displays $\partial\lambda_i/\partial\bar f_\ell$} and recolors every bus \edit{$i$} by the sign and magnitude of that response, both computed through~\eqref{eq:sens}. Fig.~\ref{fig:teaser} shows this second view on Texas7k: a binding line selected, every bus shaded green or purple by its rating sensitivity, and the rating slider at a committed $+10$~MW edit. During a drag the shading tracks the \edit{first-order} preview on every frame; releasing commits the exact re-solve, and the panel reports the prediction ($-15.9$~\$/h) beside the exact change ($-14.9$~\$/h).

\subsection{Distribution studies and agent tools}
Visualization capabilities for distribution networks currently include individual terminal voltages, currents, source powers, and \edit{voltages relative to the neutral} where a neutral is declared. Equipment inspection exposes the underlying bus, line, load, and transformer attributes alongside results. Retained multiconductor sessions allow repeated active/reactive load edits while reusing the network factorization. These updates are full nonlinear PF solves rather than sensitivity-based previews.

The \texttt{@tellegen/webmcp} package exposes structured tools to browsers implementing the WebMCP API~\cite{webmcp2026}. An agent can inspect the active case, query equipment and solved quantities, request multiconductor PF, or conduct supported transmission sensitivity and capacity experiments. A capacity proposal records its budget, trial edits, predicted changes, and re-solved results. Applying it checks the case and revision and requires approval in the interface. The experiment journal can be exported, and studies retain their inputs and solved states. Manual interaction remains available without WebMCP.

\section{Implications for Digital Transformation}\label{sec:discussion}

\subsection{Access through an executable study}
Democratizing access to power system studies is a contribution of this work. A recipient \edit{needs only} a compatible browser to run and understand a power system study. Educators can publish a network with a few meaningful controls; researchers can attach an executable example to a paper; engineers can share an interface to customers while allowing them to keep their sensitive case files local. These uses build on the access already provided by open source solvers by reducing the work required to run a particular study.

Currently, our numerical evaluations have established feasibility and realism of the \texttt{tellegen} studies, but it remains open to measure educational and organizational impact. Future work should study impacts on the classroom and users. Browser execution also shifts memory and computation to the recipient's device: a large network that runs acceptably in our benchmarks may be unsuitable for mobile users. Initial application downloads, map services, and agent connectivity are separate from local solver execution. Broader access still requires understandable defaults, accessible interfaces, suitable datasets, and clear interpretation of approximation and failure.

\subsection{Lessons for numerical software delivery}
Some of our design choices transfer beyond power systems. First, separating a fast preview of the change in a solution to an optimization problem from a completed solve keeps parameter exploration responsive. Second, running the numerical engine in a compiled worker allows the interface to remain usable during longer solves. These choices are useful for any scientific application that must support parametric edits to costly in-process optimization problems.

Critically, a local-first architecture dramatically changes the scalability. Static delivery can distribute the solver, leaving its execution cost with each client. The operator of a scientific application no longer needs a solver service to process private case files.  Large batch, multi-period, and security-constrained studies may still justify native or remote compute; the browser workflow does not require that every study fit on every device.

\subsection{Agent interaction and implementation experience}
The experimental WebMCP interface provided with \texttt{tellegen} gives an agent explicit operations over the visible study. An agent and a human can collaboratively inspect equipment, query computed quantities, preview an edit, run a bounded experiment, and propose planning decisions. The proposal workflow uses sensitivities to select trials and re-solves each candidate before reporting its result. Applying a planning proposal requires a matching case revision and explicit approval from a human, with clear records of how the plan was discovered. This architecture has potential for making AI-native interaction with power system software a reproducible use of the numerical tool rather than a wholesale replacement for its solver.

Agentic coding tools assisted implementation of the solver engine, sensitivity analysis, and browser renderer. Distribution support benefited from independent OpenDSS comparison benchmarks. This experience motivates keeping reference power grid test cases a part of the agentic development process, especially when \edit{working with} unfamiliar languages or numerical models. \edit{This also suggests there may be benefits to using established software to independently check the outputs of generated implementations.}

\subsection{Ethical considerations}
Keeping case parsing and solving on the user's device was an intentional  architectural choice. By designing our framework such that all aspects of a study can happen locally, we have produced the first web-based power system analysis tool that transmits zero CEII. Supporting broader ethical considerations around CEII was a fundamental motivation for the creation of this framework.

The validation behind these results used synthetic transmission grids and public distribution datasets. These approximate, but may not reflect, the utility network models operational use would require. Therefore, the evidence reported supports research and teaching, while operational decisions still require validation against applicable equipment models and operating requirements.

\section{Related Work}\label{sec:related}
MATPOWER~\cite{zimmerman2011matpower}, pandapower~\cite{thurner2018pandapower}, and PowerModels.jl~\cite{coffrin2018powermodels} provide open implementations for power system analysis and optimization. OpenDSS~\cite{dugan2011opendss} and PowerModelsDistribution.jl~\cite{pmd} address detailed unbalanced distribution models. \edit{\texttt{tellegen}} uses established formulations and these reference implementations to evaluate its browser delivery and interaction model.

PowerPlots.jl~\cite{rhodes2025powerplots} supports power network visualization and interactive exploration through a Julia workflow. GridSuite, built on PowSyBl~\cite{powsybl2026}, provides web applications for grid studies and is used in production at RTE. Electrisim~\cite{electrisim2026} provides browser and desktop interfaces to engines including pandapower and OpenDSS. \edit{\texttt{tellegen}} combines local file ingestion, a compiled client-side solver, and sensitivity previews tied to the same editable study.

GPU-oriented tools address a complementary requirement. ExaModelsPower.jl~\cite{johnson2025exapower} constructs nonlinear AC OPF models for GPU solvers, including multi-period and security-constrained studies. \edit{\texttt{tellegen}}'s browser workflow emphasizes distributing an interactive study to an individual user.

Sensitivity analysis predates these software architectures. Tellegen's theorem~\cite{tellegen1952}, which relates to adjoint network methods~\cite{director1969adjoint}, provides the project's historical namesake. Modern differentiable optimization layers~\cite{amos2017optnet,agrawal2019cvxpylayers} likewise obtain derivatives from solved optimality conditions. Here those derivatives provide an interactive preview and parameter modification experiments. WebMCP~\cite{webmcp2026} supplies an API for exposing application tools to agents in the browser; it is distinct from the client--server Model Context Protocol~\cite{mcp2025}. Agent interfaces for power system tools are also discussed by Zhang and Xie in~\cite{zhang2025poweragent}.

\section{Conclusion and Future Work}\label{sec:conclusion}
\texttt{tellegen} brings interactive power systems analysis and visualization to the web browser. A low-level data model and numerical engine \edit{support} local browser calculations, enabling fast, intuitive, and secure interaction with power system studies entirely on the user's machine. The evaluation compares transmission objectives with PGLib baselines, records native and wasm timings, and checks distribution voltages and currents against OpenDSS. Together, these results support distributing useful numerical studies through a web link without uploading private cases.

Validation against solvers used in industrial planning is a further direction. \texttt{tellegen} already parses PSS/E \texttt{.raw} cases, which makes comparison against PSS/E a concrete near-term target, extending validation from objectives alone to bus voltages, flows, and dispatch. Such comparison ultimately calls for the utility network models operational use would require, which are designated CEII. The local execution model of \texttt{tellegen} may ease that access, since a case can be validated on the user's own device without a server-side copy, though it does not reduce the model's underlying sensitivity.

Full nonlinear AC OPF remains future work. Its browser implementation requires a suitable nonlinear solver and numerical validation; WebAssembly itself does not rule out such a solver. Other extensions include distribution optimization controls, state estimation, and teaching materials evaluated with students.

The repository, documentation, and hosted application are available at \repourl\ and \demourl.

\section*{Disclaimer}
The views expressed in this paper are
the opinion of the authors and do not reflect the views of PJM
Interconnection, L.L.C. or its Board of Managers, of which
Le Xie is a member.

\begin{acks}
Generative AI tools assisted software development and manuscript revision, including prose editing and source and reference checks. The authors are responsible for the technical content and reported results.
\end{acks}

\bibliographystyle{ACM-Reference-Format}
\bibliography{refs}
\end{document}